\documentclass[
reprint,superscriptaddress,nofootinbib,aps,pra,twocolumn]{revtex4-1}

\usepackage{mathtools}
\usepackage[caption=false]{subfig}
\usepackage{tikz-cd}
\usepackage{multirow}
\usepackage{url}
\usepackage{hyperref}
\usepackage{cancel}
\usepackage{booktabs}
\usepackage{enumitem}
\usepackage{dsfont}
\usepackage{soul} 
\usepackage{pifont} 
\usepackage{amsmath} 
\usepackage{amssymb} 
\usepackage{amsfonts} 

\newcommand{\ket}[1]{\ensuremath{\left|\right.\!{#1}\!\left.\right\rangle}}

\newcommand{\bra}[1]{\ensuremath{\left\langle\right.\!{#1}\!\left.\right|}}
\newcommand{\braket}[2]{\ensuremath{\langle{#1}|{#2}\rangle}}
\newcommand{\ketbra}[2]{\ensuremath{|{#1}\rangle\!\langle{#2}|}}
\newcommand{\brakket}[3]{\ensuremath{\langle{#1}|{#2}|{#3}\rangle}}

\newcommand{\intsum}{\mathclap{\displaystyle\int}\mathclap{\textstyle\sum}}

\begin{document}

\title{The time of arrival problem in the Page-Wootters formalism}
\author{Niyusha Hosseini}
\email{niyusha.hosseini@tuwien.ac.at} 
\affiliation{Atominstitut, Technische Universit{\"a}t Wien, 1020 Vienna, Austria}

\author{Maximilian P.~E. Lock}
\email{maximilian.paul.lock@tuwien.ac.at} 
\affiliation{Atominstitut, Technische Universit{\"a}t Wien, 1020 Vienna, Austria}
\affiliation{Institute for Quantum Optics and Quantum Information - IQOQI Vienna, Austrian Academy of Sciences, Boltzmanngasse 3, 1090 Vienna, Austria}

\date{\today}

\begin{abstract}
The time-of-arrival problem asks for the probability distribution for when a quantum particle reaches a specified location. It has been the subject of decades of debate, exemplifying the lack of a self-adjoint time observable in quantum theory. In the Page–Wootters framework, time is a relational quantity, emerging from correlations between a system and a clock induced by a global Hamiltonian constraint. We construct a time-of-arrival distribution by inverting the Page-Wootters approach, asking what time a clock reads given that the particle arrives at some fixed position. The result coincides with a common approach to the time-of-arrival problem, suggesting a potential relational interpretation of the latter. In addition to providing a relational description of the time-of-arrival problem, this gives an application of the Page-Wootters formalism to a concrete physical problem, and reveals some complications with its canonical interpretation as a theory of conditional probabilities.

\end{abstract}

\maketitle


\section{Introduction}
\label{sec:Intro}

The treatment of time in the theory of quantum mechanics has been the subject of much debate, mainly due to the difficulty associating time with a self-adjoint operator in the canonical formalism~\cite{TimeInQuantumBook}. A concrete example where this problem becomes operational is the \emph{time-of-arrival} (ToA) problem: given an initial wave packet, how should one define the probability distribution for the time at which a particle reaches a specified location? 

In the decades since the ToA problem was first identified~\cite{AharonovBohm1961,Allcock1969,allcock1969time}, many approaches have been developed, but a consensus is still lacking. The semiclassical solution weights the trajectories of a classical particle according to the 
momentum probability distribution determined by the wave function. An arrival time distribution can also be determined from the probability current (``quantum flux'') through a surface~\cite{Allcock1969,leavens1998time,halliwell2009quantum,das2019arrival}. In an approach introduced by Kijowski~\cite{kijowski1974time}, an arrival time distribution is derived from a series of physically-motivated axioms, with the result corresponding to a covariant Positive Operator-Valued Measure (POVM)~\cite{werner1986screen}. Operator-based proposals such as the Aharonov-Bohm time operator can be recovered within this scheme~\cite{AharonovBohm1961,giannitrapani1997positive,grot1996time,egusquiza1999free}, for instance as the POVM's first-moment operator~\cite{muga1998space}. This is sometimes called the ``standard'' approach due to the variety of equivalent formulations~\cite{grot1996time,delgado1997arrival,egusquiza2002standard,leavens2002standard}. A fourth approach models arrival as a detection process via explicit system–detector dynamics~\cite{allcock1969time,aharonov1998measurement,halliwell1999arrival,yearsley2010quantum,goldstein2024arrival,reddiger2026solution}. Each route carries specific difficulties such as regularisation choices, negative contributions, and momentum-support restrictions~\cite{MugaReview2008}.

Relational approaches to time in quantum theory originate as a potential solution to the problem of time in quantum gravity~\cite{dewitt1967quantum,Rovelli1990,rovelli1991time,kuchar2011time,Isham1993,rovelliQuantumGravity2004,thiemann2008modern}. Dynamics are treated as emerging from the relationship between a system of interest and a clock, with the clock thus acting as a quantum reference frame~\cite{loveridge2018symmetry,hohn2019switching,loveridge2019relative, giacomini2019quantum,hohn2020switch}. Instead of its usual role as an external parameter, time is then understood as a property of the clock, with respect to which the system of interest should be described. The Page-Wootters formalism~\cite{PageWootters1983, woottersTimeReplacedQuantum1984,pageClockTimeEntropy1994} is one such relational approach (see~\cite{hohn2021trinity,hohn2021equivalence,chataignier2026relational} for a discussion of its equivalence to other approaches). It entails the construction of states of the system of interest relative to the clock. Starting with solutions to a Hamiltonian constraint (i.e.~the Wheeler DeWitt equation in the context of quantum gravity~\cite{dewitt1967quantum}), which encodes the invariance of a theory under reparameterisations of time, one obtains the states of the system ``conditioned'' on the clock reading a given time, which then satisfy the usual Schr\"{o}dinger equation.

The ToA problem has been studied from a relational perspective in a few inequivalent ways. In~\cite{gambini2022solution}, the arrival time is described via conditional probabilities with respect to a clock, using a formalism where finite clock energy leads to nonunitary effective evolution and thus decoherence~\cite{gambini2007fundamental}. In~\cite{hohn2020switch}, on the other hand, the regularisation of the Aharonov–Bohm operator from~\cite{grot1996time} results in equivalent regularised self-adjoint ToA operators in both Dirac and quantum-deparametrised relational pictures. Finally, in the ``quantum clock'' proposal~\cite{MacconeSacha2020,roncallo2023does,pintocouto2026first}, one restricts the relational description to a finite range of time, whose width is interpreted as a regularisation parameter, to obtain a conditional probability of a clock time given a particular particle position. The role of this regularisation parameter (required for a finite conditional probability), and the dependence of the ToA probability thereon, has been debated~\cite{cavendish2024physical,maccone2025quantum}.

In this work, we introduce a relational solution to the ToA problem based on the Page-Wootters formalism mentioned above. This gives a novel approach to the problem, while providing a tangible example to which the rather abstract Page-Wootters formalism can be applied. We show how to obtain a well-defined relational ToA probability distribution by inverting the usual Page-Wootters approach to obtain states of a clock described relative to the particle's position, finding that the resulting distribution coincides with the one derived by Kijowski~\cite{kijowski1974time}. We find that the structural features of the formalism, in particular the direct-sum decomposition of the physical Hilbert space implied by the form of the Hamiltonian constraint, prohibit interference between momentum branches, and that this feature is independent of the choice of clock observable. This example reveals certain complications in the usual interpretation of the Page-Wootters formalism in terms of conditional probabilities, as we discuss in detail.

The paper is organised as follows. In Sec.~\ref{sec:Page-Wootters} we review the Page-Wootters framework for relational dynamics, emphasising certain mathematical details which are often overlooked, but which are relevant for the problem at hand, and then describing some peculiarities that arise when the system of interest is a free particle. Sec.~\ref{sTOAinPage-Wootters} develops the construction of a relational ToA probability distribution, examining how the constraint demands the separate consideration of momentum sectors, and how to obtain conditional states of the clock which translate appropriately under translations of the particle position. As noted above, the resulting probability distribution coincides with the one derived by Kijowski. Motivated by this result, Sec.~\ref{sec:necessity_of_conditional_observables} critically analyses the usual interpretation of Page-Wootters as a theory of conditional probabilities. We discuss the implications of our work and its relation to the existing literature in Sec.~\ref{sDisc}, before concluding in Sec.~\ref{sec:conclusion}. Throughout this work, we set $\hbar=1$ and adopt the convention that tensor products between states are implicit unless otherwise indicated.

\section{The Page--Wootters Formalism}
\label{sec:Page-Wootters}

\subsection{``Timeless'' physical states}
\label{sec:H_phys}

We first briefly review some elements of the quantisation of a constrained system~\cite{marolf1995refined,hartle1997comparing,marolf2000group,thiemann2008modern}, which forms the basis for the Page–Wootters formalism. Let $\mathcal{H}_{\mathrm{kin}}$ denote a kinematical Hilbert space, and  $\hat{C}$ a Hermitian operator defined thereupon, identified with the Hamiltonian of some symmetry-constrained system. The \textit{physical states} are those satisfying the Hamiltonian constraint, known in the context of quantum gravity as the Wheeler–DeWitt equation~\cite{dewitt1967quantum}:
\begin{equation} \label{eHamConstraint}
\hat{C} |\psi_{\text{phys}}\rangle = 0,
\end{equation}
which encodes a one-parameter gauge symmetry:
\begin{equation}
e^{i \hat{C} r} |\psi_{\text{phys}}\rangle = |\psi_{\text{phys}}\rangle \quad \forall\, r\in\mathds{R} .
\end{equation}
The physical states are then invariant under ``evolution'' generated by this Hamiltonian. On the other hand, any two kinematical states related to each other by an element of this group, say $|\psi'_{\text{kin}}\rangle=e^{i \hat{C} r} |\psi_{\text{kin}}\rangle$, contain the same physical information, being related by a gauge transformation. Thus, the kinematical space ``over-counts'' the number of states and must be reduced to a space containing only physically meaningful, i.e. gauge-invariant, information.

The set of solutions to Eq.~\eqref{eHamConstraint} constitute a physical Hilbert space $\mathcal{H}_{\text{phys}}$ of states satisfying the requisite symmetry, where the inner product between two physical states is given by
\begin{equation}
\label{eq:abstract_phys_IP}
\braket{\psi_{\mathrm{phys}}}{\phi_{\mathrm{phys}}}_{\mathrm{phys}}
\coloneqq
\big\langle\psi_{\text{kin}}\big|\delta(\hat{C})\big|\phi_{\text{kin}}\big\rangle_{\text{kin}} ,
\end{equation}
where
\begin{equation}
\label{eq:phys_projector}
\delta(\hat{C})
\coloneqq
\frac{1}{2\pi}\int_{\mathbb{R}} dr\; e^{i r \hat{C}} 
\end{equation}
is the (in general improper) projector onto the physical Hilbert space $\delta(\hat{C}):\mathcal{H}_{\mathrm{kin}}\to\mathcal{H}_{\text{phys}}$, and where $|\psi_{\text{kin}}\rangle$ (respectively $|\phi_{\text{kin}}\rangle$) is any member of the equivalence class of states in $\mathcal{H}_{\mathrm{kin}}$ which map to $|\psi_{\text{phys}}\rangle$ (respectively $|\phi_{\text{phys}}\rangle$) under the action of $\delta(\hat{C})$. In this way, the constraint both selects the physical states and induces the inner product on the resulting physical Hilbert space. Defining the inner product as in Eq.~\eqref{eq:abstract_phys_IP} prevents the divergences that would otherwise occur, since solutions to Eq.~\eqref{eHamConstraint} are not in general normalisable in $\mathcal{H}_{\mathrm{kin}}$.

Thus, unlike in ``standard'' quantum mechanics, where evolution occurs with respect to an external time parameter, the physical states here are ``frozen'' in the sense that they do not evolve under the group generated by $\hat C$. This is the so-called problem of time, arising due to the Hamiltonian constraint~\cite{kuchar2011time,Isham1993}. The question is then how evolution can emerge from such timeless physical states.

\subsection{Dynamics relative to a quantum clock}
\label{sec:Page-Wootters_A}

As noted above, the Page-Wootters proposal~\cite{PageWootters1983, woottersTimeReplacedQuantum1984,pageClockTimeEntropy1994} recovers dynamics relationally, by describing the system relative to an internal clock degree of freedom rather than an external parameter, allowing a meaningful notion of time evolution to emerge from otherwise timeless states. Accordingly, one assumes a kinematical factorisation into a clock and a system,
\begin{equation}
\mathcal{H}_{\mathrm{kin}} \simeq \mathcal{H}_{\mathrm{C}}\otimes\mathcal{H}_{\mathrm{S}}.
\end{equation}
Assuming no interactions between clock and system, the constraint Hamiltonian takes the form:
\begin{equation}
\hat{C} = \hat{H}_C \otimes \mathds{1}_S + \mathds{1}_C \otimes \hat{H}_S.
\end{equation}
We assume that the clock Hamiltonian $\hat{H}_C$ has a nondegenerate spectrum, so that its energy eigenstates $|\varepsilon\rangle_C$ are uniquely labelled by the clock energies $\varepsilon$. We denote the eigenstates of $\hat{H}_S$ by $|E,\sigma_E\rangle_S$, where $\sigma_E$ labels the possible degeneracies at system energy $E$. The tensor-product basis $\{|\varepsilon\rangle_C |E,\sigma_E\rangle_S \}_{\varepsilon,E,\sigma_E}$ can then be used to describe states in $\mathcal{H}_{\text{kin}}$.

We now require a suitable observable on $\mathcal{H}_{\mathrm{C}}$ to represent the clock time. This is provided by a POVM which is covariant with respect to the clock Hamiltonian~\cite{holevoProbabilisticStatisticalAspects1982,buschOperationalQuantumPhysics}, constructed in the following manner. We first seek a family of ``time states'' on $\mathcal{H}_{\mathrm{C}}$, via the covariance condition:
\begin{equation} \label{eCovStates}
|t + t'\rangle \equiv e^{-i \hat{H}_C t'} |t\rangle, 
\end{equation}
which allows us to define a covariant POVM on $\mathcal{H}_{C}$ with elements $\{E(t) \coloneqq \ketbra{t}{t}\}_{t \in \mathds{T}}$, subject to the normalisation condition:
\begin{equation} \label{ePOVMNorm}
N \int_\mathds{T} dt \, E(t) = \mathds{1},
\end{equation}
where $\mathds{T}$ and $N$ are determined by $\hat{H}_{\mathrm{C}}$ (see, e.g.,~Sec.~III of~\cite{hohn2021trinity}). Hence, $t \in \mathds{T}$ is interpreted as a probabilistic measurement outcome labelling the elements of this POVM. Eqs.~\eqref{eCovStates} and~\eqref{ePOVMNorm} imply that the time states must satisfy
\begin{equation}
|t\rangle = \, \,\,\,{{\intsum}_{\text{spec}(\hat{H}_C)}} d\varepsilon \, e^{i f(\varepsilon)} e^{-i \varepsilon t} |\varepsilon\rangle,
\end{equation}
where $f(\varepsilon)$ is an arbitrary real function, which we henceforth choose as $f(\varepsilon)=0$.

Now, given a physical state $|\psi_\text{phys}\rangle$, which is described in terms of the kinematical structures $C$ and $S$ (the ``perspective-neutral'' description~\cite{hohn2019switching,de2021perspective}), we can use the time states defined above to obtain an equivalent, ``reduced'', representation associated with the system $S$ subject to the clock reading the time $t$. In particular, for each $t$, one defines the Page-Wootters reduction map:
\begin{equation}
    \mathcal{R}_{C}(t)\coloneqq \langle t|_{C} \otimes \mathds{1}_S ,
\end{equation}
and the corresponding conditional system state associated with a physical state $|\psi_\text{phys}\rangle$:
\begin{equation} \label{eReducedStates}
|\psi_{S\vert C}(t)\rangle :=  \mathcal{R}_{C}(t) |\psi_\text{phys}\rangle.
\end{equation}
The family of states $\{|\psi_{\mathrm{S|C}}(t)\rangle\}_{t\in\mathds{T}}$ are elements of the reduced (conditional) Hilbert space $\mathcal{H}_{\mathrm{S|C}}$~\cite{hohn2021trinity}, where the latter is defined by the image of $\mathcal{H}_\mathrm{phys}$ under $\mathcal{R}_{C}(t)$, and is independent of $t$. To obtain an equivalent physical description, the reduction must preserve the physical norm, i.e.
\begin{equation}
\langle\psi_{\mathrm{S|C}}(t)|\psi_{\mathrm{S|C}}(t)\rangle = \langle\psi_{\mathrm{phys}}|\psi_{\mathrm{phys}}\rangle_{\mathrm{phys}} \quad \forall \, t .
\end{equation}
This is indeed the case, and in fact gives an alternative (and equivalent~\cite{hohn2021trinity,hohn2021equivalence,chataignier2026relational}) definition of the physical inner product~\cite{smith2019quantizing}. Thus the reduction is an isometry, i.e.~$\mathcal{H}_{\mathrm{S|C}}\simeq\mathcal{H}_\mathrm{phys}$. Using the normalisation condition of the POVM, Eq.~\eqref{ePOVMNorm}, the physical state $|\psi_\text{phys}\rangle$ can then be written as a ``history'' state:
\begin{equation}
|\psi_\text{phys}\rangle = N \int_{\mathbb{T}} dt \, |t\rangle_C  |\psi_{S\vert C}(t)\rangle,
\end{equation}
where the integral ranges over all clock readings $t \in \mathbb{T}$, and each $|\psi_{S\vert C}(t)\rangle$ represents the state of the system conditioned on the clock reading $t$, and satisfies the Schr\"{o}dinger equation with respect to this parameter~\cite{PageWootters1983, woottersTimeReplacedQuantum1984}. For this reason, the Page-Wootters formalism is sometimes called the ``conditional probability interpretation'' of dynamics, a point which we discuss in detail in Sec.~\ref{sec:necessity_of_conditional_observables}.

Note that despite its association with the system, in general $\mathcal{H}_{\mathrm{S|C}}\not\simeq \mathcal{H}_{\mathrm{S}}$~\cite{hohn2021trinity,hohn2021equivalence,chataignier2026relational}, as $\hat{H}_C$ determines what subset (if any) of the eigenstates of $\hat{H}_S$ is compatible with the constraint equation. If, for example, we have $\text{spec}(\hat{H}_S)\subseteq\text{spec}(-\hat{H}_C)$ then $\mathcal{H}_{\mathrm{S|C}}\simeq \mathcal{H}_{\mathrm{S}}$. The requirement of compatibility with Eq.~\eqref{eHamConstraint} applies not only to the definition of the states of the system relative to the clock, but also to the definition of operators on the conditional Hilbert space. In particular, given some kinematical operator $\hat{O}_S$ on $\mathcal{H}_{\mathrm{S}}$, the corresponding operator $\hat{O}_{S|C}$ on $\mathcal{H}_\mathrm{S|C}$ is defined by restricting both the domain and range of $\hat{O}_S$ to states that are compatible with Eq.~\eqref{eHamConstraint}. With this notation in hand, the reduced states defined in Eq.~\eqref{eReducedStates} evolve on $\mathcal{H}_{\mathrm{S|C}}$ with respect to the clock time $t$ according to the Schr\"{o}dinger equation: $i\partial_{t}|\psi_{S\vert C}(t)\rangle=\hat{H}_{S|C}|\psi_{S\vert C}(t)\rangle$.

In the case where the clock Hamiltonian $\hat{H}_C$ is degenerate, the physical Hilbert space decomposes into a direct sum of degeneracy-sectors, and the reduction maps $\mathcal{R}_{C}(t)$ must be defined sector-wise~\cite{hohn2021equivalence}. In that case these maps are partial isometries, and the physical inner product is recovered after summing the corresponding reduced inner products over sectors. We will return to these features in Sec.~\ref{sTOAinPage-Wootters}.

\subsection{Describing a free particle} \label{secFreeParticlePage-Wootters}

We now specialise to the case where the system of interest is a free particle, i.e.~the setting of the ToA problem. Given $\hat{H}_S = \frac{\hat{p}^2}{2m}$, the constraint operator takes the form
\begin{equation} \label{eHamConstFreeP}
\hat C=\hat{H}_C \otimes \mathds{1}_S + \mathds{1}_C \otimes\frac{\hat p^{2}}{2m},
\end{equation}
where $\hat p$ is the particle's momentum operator. Now, exchanging the roles of clock and system in Eq.~\eqref{eHamConstFreeP} gives an equivalent constraint operator to the one considered in~\cite{hohn2021equivalence}. One may then follow the same procedure (see also~\cite{hohn2019switching,hohn2020switch}) to describe the physical Hilbert space, factorising the constraint according to ${\hat C =\hat C_+ \hat C_-}$, with
\begin{equation}
\hat C_\sigma \coloneqq \frac{\hat p}{\sqrt{2m}} - \sigma \sqrt{-\hat H_C} \,,
\end{equation}
where $[\hat C_+ ,\hat C_-]=0$. The cases $\sigma=+$ and $\sigma=-$ correspond to the positive- and negative-momentum solutions to the constraint equation, Eq.~\eqref{eHamConstraint}. Note that we use a different sign convention for $\sigma$ here compared to~\cite{hohn2019switching,hohn2020switch,hohn2021equivalence}. The projector $\delta(\hat C)$ defined in Eq.~\eqref{eq:phys_projector} can then be decomposed into two branches,
\begin{equation} \label{eSumSeparationProjector}
\delta(\hat C) = \frac{1}{2\sqrt{-\hat H_C}} \sum_{\sigma = \pm} \delta(\hat C_\sigma),
\end{equation}
illustrating how Eq.~\eqref{eHamConstFreeP} induces a direct-sum decomposition of the physical Hilbert space:
\begin{equation}
\mathcal{H}_{\mathrm{phys}}=\bigoplus_{\sigma=\pm}\mathcal{H}^{\sigma}_{\mathrm{phys}},
\end{equation}
where $\mathcal{H}^{\sigma}_{\mathrm{phys}}$ consists of the solutions to $\hat C_{\sigma}\ket{\psi_{\text{phys}}}=0$.

In order to obtain a relational description of a free particle which behaves appropriately, we must have ${\hat{H}_{S|C}=\hat{p}^2 /2m}$, or in other words that the system Hamiltonian in the relational description behaves just as $\hat{H}_S$ on the kinematical factor. Considering the zero-eigenvalues of the constraint operator in Eq.~\eqref{eHamConstFreeP} in terms of the eigenvalues of $\hat{H}_C$ and $\hat{H}_S$, we can see how $\hat{H}_C$ is constrained by this demand. Specifically, noting that
\begin{equation*}
  \text{spec}(\hat{H}_{S|C}) = \{ E \in \text{spec}(\hat{H}_S) \mid -E \in \text{spec}(\hat{H}_C) \},
\end{equation*}
and moreover that $\text{spec}(\hat H_S)=\mathbb R_+$, we see that the requirement $\text{spec}(\hat{H}_{S|C})=\text{spec}(\hat H_S)$ implies that $\mathbb R_-\subseteq\text{spec}(\hat H_C)$. This in turn fixes the normalisation constant in Eq.~\eqref{ePOVMNorm} as $N=1/2\pi$, as well as determining the set of clock times to be $\mathbb{T}=\mathds{R}$ (cf.~the regularisation parameter in~\cite{MacconeSacha2020,roncallo2023does,pintocouto2026first}). In this case, we also have $\mathcal H_{S|C}\simeq \mathcal H_S$. We hereafter assume $\mathbb R_-\subseteq\text{spec}(\hat H_C)$, permitting a relational picture of a free particle.

Let us write an arbitrary element of the kinematical space $\mathcal{H}_{\mathrm{kin}}$ in the tensor-product basis of the clock Hamiltonian and the particle momentum,
\begin{equation}
    \ket{\psi_{\text{kin}}}
=\int_{\mathds{R}} dp \int_{\mathrm{spec} (\hat{H}_{C})} d\varepsilon\;
\;
\psi_{\mathrm{kin}}\!(\varepsilon, p) 
\ket{\varepsilon}_C  \ket{p}_S .
\end{equation}
On $\mathcal{H}_{\mathrm{phys}}$, each clock energy $\varepsilon$ corresponds to two momenta $p_\sigma(\varepsilon)=\sigma\sqrt{-2m\varepsilon}$,  and the corresponding physical state can then be written as
\begin{equation} \label{ePhysStateFreeP}
\ket{\psi_{\text{phys}}}
=\sum_{\sigma=\pm}\int_{\mathds{R}_-} d\varepsilon\;
\frac{\sigma m}{p_\sigma(\varepsilon)}
\psi_{\mathrm{kin}}\!\big(\varepsilon, p_\sigma(\varepsilon)\big)
\ket{\varepsilon}_C  \ket{p_\sigma(\varepsilon)}_S .
\end{equation}
Inserting Eq.~\eqref{eSumSeparationProjector} into the definition of the physical inner product, i.e.~Eq.~\eqref{eq:abstract_phys_IP}, yields
\begin{multline}
\braket{\psi_{\mathrm{phys}}}{\phi_{\mathrm{phys}}}_{\mathrm{phys}}
= \\
\int_{\sigma_{\mathrm{C}}} d\varepsilon\;
\sqrt{-\frac{m}{2\varepsilon}}\;
\sum_{\sigma=\pm}
\psi_{\mathrm{kin}}^{*}\!\bigl(\varepsilon, p_\sigma(\varepsilon)\bigr)\, 
\phi_{\mathrm{kin}}\!\bigl(\varepsilon, p_\sigma(\varepsilon)\bigr) \, ,
\label{eq:physical_inner_product}
\end{multline}
In particular, physical expectation values are insensitive to coherence between the $\sigma=\pm$ components. In other words, the constraint enforces a superselection rule between the two sectors (see, e.g.,~\cite{bartlett2007reference,hohn2021equivalence}).

Conditioning on a clock reading $t$ in the manner described in Sec.~\ref{sec:Page-Wootters_A}, in particular Eq.~\eqref{eReducedStates}, yields the usual Page-Wootters conditional system state,
\begin{equation}
\label{eq:cond_state_time}
|\psi_{S|C}(t)\rangle
= \int_{\mathbb{R}} dp\;\psi_{\text{kin}}\!\big(\varepsilon(p),p\big)\,e^{-i\frac{p^2}{2m}t}\,|p\rangle_S ,
\end{equation}
where $\varepsilon(p)\coloneqq-p^{2}/(2m)$.
Thus $\psi_{\mathrm{kin}}\!\bigl(\varepsilon(p),p\bigr)$ plays the role of the ``initial'' momentum-space wavefunction, i.e.
\begin{equation} \label{eInitialWavefunction}
    \psi_0(p)\coloneqq\braket{p}{\psi_{S|C}(0)} = \psi_{\mathrm{kin}}\!\bigl(\varepsilon(p),p\bigr) ,
\end{equation}
illustrating how the choice of the initial state in ``usual'' quantum theory is replaced by the choice of the physical state in the Page-Wootters formalism.


\section{Posing the time-of-arrival problem in the Page-Wootters formalism} \label{sTOAinPage-Wootters}

\subsection{The clock state conditioned on the system position} \label{sConditionPosition}

We seek a purely relational treatment of arrival times which respects the structure of the physical Hilbert space (and thus the Hamiltonian constraint). Specifically, following the Page-Wootters formalism, rather than asking for the state of the particle when the clock reads a certain time, \textit{we wish to describe the state of the clock when the particle is at some definite position, say $x_{0}$}. The probability distribution for the ToA at $x_{0}$ can then be obtained from this relational clock state via the clock-time POVM described in Sec.~\ref{sec:Page-Wootters_A}.

Let us first consider the naive use of position eigenstates $\ket{x}_{S}$ to define a reduction map ${\tilde{\mathcal{R}}(x_{0})\coloneqq}\mathds{1}_C \otimes \bra{x_{0}}_S$ acting on $\mathcal{H}_{\mathrm{phys}}$. By analogy with the usual Page-Wootters procedure, one may then attempt to interpret $\tilde{\mathcal{R}}(x_{0})\ket{\psi_{\text{phys}}}$ as the relational state of the clock when the particle is at $x_{0}$. However, from Eq.~\eqref{ePhysStateFreeP}, one finds that the resulting state coherently combines the different superselection sectors in $\mathcal{H}_\mathrm{phys}$, encoding symmetry-forbidden phase information between sectors. Crucially, $\tilde{\mathcal{R}}(x_{0})$ is not invertible, and the relational clock states, and therefore the arrival time probability distribution derived therefrom, cannot be normalised in a state-independent manner, and we must therefore find another solution.

In order to obtain a normalised clock observable relative to the system position, a reduction map, when acting on a normalised physical state $\ket{\psi_{\text{phys}}}$, must give a corresponding normalisation of the relational clock states. In other words, there must be isometry between $\mathcal{H}_{\mathrm{phys}}$ and the space of relational states of the clock. However, as shown in Appendix~\ref{sSectorwiseReduction}, isometry is only possible per superselection sector (i.e. partial isometry), thus implying that a general reduction map with respect to some property of $\mathcal{H}_{\mathrm{S}}$ must be defined per superselection sector, just as in the case of relational dynamics with respect to a quadratic Hamiltonian~\cite{hohn2021equivalence}. We now consider the general form that such a map must take, and then show how the remaining freedom is fixed by the appropriate spatial translation property, namely covariance.

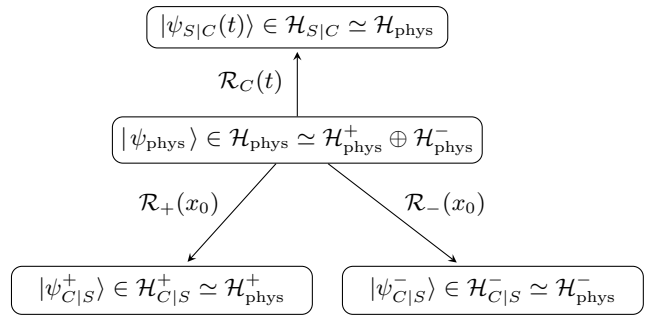
\begin{figure}[t]
  \centering
  \begin{tikzpicture}[
      >=stealth,
      every node/.style={font=\small},
      box/.style={
        draw,
        rounded corners,
        inner sep=3pt,
        minimum width=4.0cm
      }
    ]

    \node[box] (PsiSC) at (0,1.3) {$|\psi_{S|C}(t)\rangle \in \mathcal{H}_{S|C}\simeq\mathcal{H}_{\mathrm{phys}}$};

    \node[box] (Hphys) at (0,-0.2)
      {$\ket{\psi_\mathrm{phys}}\in\mathcal{H}_{\mathrm{phys}}\simeq\mathcal{H}_\mathrm{phys}^{+}\oplus\mathcal{H}_\mathrm{phys}^{-}$};

    \node[box] (plus) at (-1.8,-2.2)
      {$|\psi_{C|S}^{+}\rangle \in \mathcal{H}_{C|S}^{+} \simeq \mathcal{H}_\mathrm{phys}^{+}$};

    \node[box] (minus) at (2.6,-2.2)
      {$|\psi_{C|S}^{-}\rangle \in \mathcal{H}_{C|S}^{-} \simeq \mathcal{H}_\mathrm{phys}^{-}$};

    \draw[->,shorten >=1.3pt,shorten <=0pt]
      (Hphys) -- node[left=2pt] {$\mathcal{R}_{C}(t)$} (PsiSC);

    \draw[->,shorten >=2pt,shorten <=0pt]
      (Hphys) -- node[above left=0.6pt, yshift=-4.5pt] {$\mathcal{R}_{+}(x_{0})$} (plus);

    \draw[->,shorten >=2pt,shorten <=0pt]
      (Hphys) -- node[above right=0.6pt, yshift=-4.5pt] {$\mathcal{R}_{-}(x_{0})$} (minus);

  \end{tikzpicture}
  \caption{Illustration of the relational Hilbert spaces associated with the clock and the particle, and the corresponding maps from the physical Hilbert space.}
  \label{fig:plus-minus-branches}
\end{figure}

Let us denote the reduction map associated with the particle being in position $x_{0}$, for a given $\sigma$-sector by  $\mathcal{R}_{\sigma}(x_{0})\coloneqq\mathds{1}_C \otimes \bra{x_{0},\sigma}_S$. As shown in Appendix~\ref{sSectorwiseReduction}, the isometry requirement discussed above implies that $\ket{x_{0},\sigma}_S$ must be of the form
\begin{equation} \label{eRequiredReductionFormMain}
\ket{x_{0},\sigma}_{S} = \int_\mathds{R} dp \, \Theta(\sigma p) \,\sqrt{\frac{\sigma p}{m}} e^{i     \varphi_{\sigma}(p , x_{0})} \ket{p}_{S}    
\end{equation}
for some real function $\varphi_{\sigma}(x_{0},p)$ of $p$, where $\Theta (p)$ denotes the Heaviside function. Now, considering the translation operator on $\mathcal{H}_{S}$, namely $\hat{T}(x)\coloneqq e^{-ix\hat{p}}$, and noting that $\mathds{1}_{C}\otimes\hat{T}(x)$ commutes with the constraint (and thus ${\mathds{1}_{C}\otimes\hat{T}(x)\ket{\psi_\mathrm{phys}}\in\mathcal{H}_\mathrm{phys}}$ $\forall\ket{\psi_\mathrm{phys}}\in\mathcal{H}_\mathrm{phys}$), we demand that
\begin{equation} \label{eTranslationProperty}
    \mathcal{R}_{\sigma}(x_{0})\big(\mathds{1}_{C}\otimes\hat{T}(x')\big) \overset{!}{=} \mathcal{R}_{\sigma}(x_{0}-x') .
\end{equation}
In other words, translating the $x$-axis by an amount $x'$, and then finding the state of the clock associated with a particle arriving at some position $x_{0}$ is the same as finding the state of the clock associated with the corresponding point before the translation, i.e.~$x_{0}-x'$. Equivalently, we require that $\hat{T}(x')\ket{x_{0},\sigma}_{S}=\ket{x_{0}+x',\sigma}_{S}$, (i.e.~covariance of the states under spatial translation) which in turn implies that, $\varphi_{\sigma}(x_{0},p)=-x_{0}p+\varphi_{\sigma}(p)~{(\mathrm{mod}~2\pi)}$, where $\varphi_{\sigma}(p)$ is an arbitrary function, which we can set to $0$ without loss of generality (cf.~the arbitrary function of energy permitted by the covariance condition with respect to time~\cite{hohn2021trinity}). The sector-wise ``position'' states are thus
\begin{equation} \label{ePositionReductonStates}
\ket{x_{0},\sigma}_{S} = \int_\mathds{R} dp \, \Theta(\sigma p) \,\sqrt{\frac{\sigma p}{m}} e^{-i  x_{0}p} \ket{p}_{S}    ,
\end{equation}
giving a partial isometry $\mathcal{R}_{\sigma}(x_{0}):\mathcal{H}_\mathrm{phys}\to\mathcal{H}_{C|S}^{\sigma}$, where $\mathcal{H}_{C|S}^{\sigma}\simeq\mathcal{H}_{C}$ and $\mathcal{H}_{C|S}^{\sigma}\simeq\mathcal{H}_\mathrm{phys}^{\sigma'}$ $\forall\, \sigma,\sigma'$.\footnote{Note that the reduction map $\mathcal{R}_{\sigma}(x_{0})$ can also be obtained by projecting from $\mathcal{H}_\mathrm{phys}$ onto a $\sigma$-sector $\mathcal{H}_\mathrm{phys}^{\sigma}$, applying $\mathds{1}_{C}\otimes\bra{x_{0}}_{S}$ and then normalising the result. Furthermore, it is a consequence of our assumption that all eigenvalues of $\hat{H}_C$ are compatible with the constraint that $\mathcal{H}_{C|S}^{\sigma}\simeq\mathcal{H}_{C}$; more generally $\mathcal{H}_{C|S}^{\sigma}\subseteq\mathcal{H}_{C}$.} The map $\mathcal{R}_{\sigma}(x_{0})$ is invertible on $\mathcal{H}_\mathrm{phys}^{\sigma}$ with inverse
\begin{equation}
    \mathcal{R}_{\sigma}(x_{0})^{-1} = \delta (\hat{C}) \big( \mathds{1}_{C}\otimes\ket{x_{0},\sigma}_{S} \big),
\end{equation}
as shown in Appendix~\ref{sSectorwiseReduction}, and in the same manner as Eq.~(42) in~\cite{hohn2021equivalence}. Then $\sum_{\sigma}\mathcal{R}_{\sigma}(x_{0})^{-1}\mathcal{R}_{\sigma}(x_{0})=\mathds{1}_\mathrm{phys}$. The relationship between the relevant Hilbert spaces is depicted in Fig.~\ref{fig:plus-minus-branches}.

We finally arrive at the state of the clock when the particle is at position $x_{0}$, and in sector $\sigma$, i.e.,
\begin{equation}
\begin{aligned} \label{eRelationalClockState}
    \ket{\psi_{C|S}^{\sigma}(x_{0})} \coloneqq  \mathcal{R}_{\sigma}(x_{0})\ket{\psi_\mathrm{phys}} 
    \equiv \big(\mathds{1}_C \otimes \bra{x_{0},\sigma}_S \big) \ket{\psi_\mathrm{phys}} \\
    = \sigma \int_{\sigma_{C|S}}  d\varepsilon  \left(-\frac{m}{2\varepsilon}\right)^{\frac{1}{4}}  \psi_{\text{kin}}(\varepsilon, p_{\sigma}(\varepsilon) ) e^{i x_{0} p_{\sigma}(\varepsilon)} \ket{\varepsilon}_C ,
\end{aligned}
\end{equation}
which is analogous to the usual Page-Wootters state in Eq.~\eqref{eq:cond_state_time}. Note that these states are not normalised on $\mathcal{H}_{C|S}^{\sigma}$, and they instead satisfy $\braket{\psi_{C|S}^{\sigma}(x_{0})}{\psi_{C|S}^{\sigma}(x_{0})}=\bra{\psi_\mathrm{phys}}{\Pi_\mathrm{phys}^{\sigma}}\ket{\psi_\mathrm{phys}}$, where $\Pi_\mathrm{phys}^{\sigma}:\mathcal{H}_\mathrm{phys}\to\mathcal{H}_\mathrm{phys}^{\sigma}$ is the projector from the physical space to the corresponding $\sigma$-sector. We then have, as a consequence of the normalisation of $\ket{\psi_\mathrm{phys}}$, that $\sum_{\sigma}\braket{\psi_{C|S}^{\sigma}(x_{0})}{\psi_{C|S}^{\sigma}(x_{0})}=1$. This, as we shall see, leads to the normalisation of the arrival time probability distribution.

\subsection{Relational time of arrival distribution} \label{sToADistibution}

Given the relational description of the clock relative to the particle position $x_{0}$, i.e.~Eq.~\eqref{eRelationalClockState}, we can compute the ToA distribution. Specifically, the probability density for the clock to read time \( t \), given that the particle is found at \( x_{0} \), is computed using the POVM element \( E(t) \) from Sec.~\ref{sec:Page-Wootters_A}, i.e.
 
\begin{equation} \label{eq:arrival-distribution-states}
    \Pr[t|x_{0}] = \sum_\sigma \brakket{\psi^\sigma_{C\vert S}(x_{0})}{\, E(t)\, }{\psi^\sigma_{C\vert S}(x_{0})}
\end{equation}
Note that the above probability distribution is normalised by construction, following from the normalisation of the conditional clock states (inherited from $\ket{\psi_{\mathrm{phys}}}$ via the physical inner product), together with the completeness of the covariant POVM, $E(t)$. Hence, no separate regularisation is required. Writing this in terms of the physical state in Eq.~\eqref{ePhysStateFreeP}, we have 
\begin{equation}
\label{eq:toa_density_full_pm}
\begin{aligned}
P[t|x_0]=& \\
 \frac{1}{2\pi}\sum_{\sigma=\pm}&
\left\vert
\int_{0}^{\infty} dp \;
\sqrt{\frac{p}{m}}\;
\psi_{\mathrm{kin}}\!\left(-\frac{p^{2}}{2m},\,\sigma p\right) 
e^{i\left(\sigma p x_0 - \frac{p^{2}}{2m}t \right)}
\right\vert^{2}.
\end{aligned}
\end{equation}
To compare with other approaches, it is convenient to express the result in terms of an ``initial'' state. 
In the Page-Wootters framework, no preferred initial time is singled out, since all clock readings are encoded in the physical state and evolution is recovered through conditioning. Then, as discussed in Sec.~\ref{secFreeParticlePage-Wootters} above, one can define a relational initial state as the state of the particle when the clock reads $t=0$, i.e.~Eq.~\eqref{eInitialWavefunction}, showing how the choice of initial condition is equivalent to the choice of physical state. The resulting probability distribution is then simply
\begin{equation} \label{eToADistributionInit}
\begin{aligned}
P[t|x_0]
&= \frac{1}{2\pi}\sum_{\sigma=\pm}
\Biggl|
\int_{0}^{\infty} dp \;
\sqrt{\frac{p}{m}}\;
\psi_0(\sigma p) \ 
e^{i\sigma p x_0- i\frac{p^{2}}{2m}t}
\Biggr|^{2},
\end{aligned}
\end{equation}
from which we see that the relational ToA distribution defined above coincides with the one derived by Kijowski~\cite{kijowski1974time} (see, e.g.,~Sec~II\hspace{0.5mm}A in~\cite{galapon2005transition} for a more readily-comparable form). Here, however, the separation into left- and right-moving sectors is not assumed, but rather derived as a necessary consequence of the symmetry constraint.

\section{Conditional probabilities and the physical meaning of the Page-Wootters formalism}
\label{sec:necessity_of_conditional_observables}

In the Page-Wootters formalism one often speaks of the probability that a system observable has some outcome \textit{conditioned on} the clock reading time $t$. Indeed, as noted above, the formalism itself is sometimes referred to as the ``conditional probability interpretation''~(e.g.~\cite{kuchar2011time}). Can the arrival time distribution in Sec.~\ref{sTOAinPage-Wootters} then be seen as the conditional probability of the clock time, when the particle is at position $x_0$? There are certain complications with this interpretation, as we now discuss, but these are in fact not unique to the problem of arrival time, instead exemplifying a broader issue. We begin with a general discussion of the relationship between physical and kinematical structures, and consider an illustrative example in the Page-Wootters formalism, before returning to the particular case of the arrival-time problem.

A conditional probability distribution, say for $A$ given some value of $B$, is determined by the joint distribution for $A$ and $B$, and the marginal distribution for $B$ (itself obtainable from the joint distribution). Considering some bipartite kinematical Hilbert space ${\mathcal{H}_\mathrm{kin}\simeq\mathcal{H}_\mathrm{A}\otimes\mathcal{H}_\mathrm{B}}$, the construction of joint and marginal probabilities, and thus the conditional probability for some observable on $\mathcal{H}_\mathrm{A}$ given the outcome of some observable on $\mathcal{H}_\mathrm{B}$ is straightforward. However, this picture is complicated by the Hamiltonian constraint, as the physical states do not factorise:~$\mathcal{H}_\mathrm{phys}\not\simeq\mathcal{H}_\mathrm{A}\otimes\mathcal{H}_\mathrm{B}$. Thus $\mathcal{H}_\mathrm{phys}$ does not simply correlate A and B,\footnote{See Sec.~VI in~\cite{hohn2021trinity} for a detailed critical discussion of this ``time from entanglement'' perspective.} but rather confounds them into one system, such that the only allowable probability distributions derived therefrom necessarily combine elements of both, even when derived from operators which are separable in kinematical terms. This is particularly prominent in the multipartite case, where tensor factorisation is clock-dependent~\cite{ali2022quantum}, so system-local observables with respect to one clock appear in general as combined system-clock observables when described with respect to a second clock (see, e.g.,~\cite{hohn2021trinity,baumann2026time}). 

To illustrate this, consider some arbitrary POVMs on $\mathcal{H}_\mathrm{A}$ and $\mathcal{H}_\mathrm{B}$ with respective elements $E_{A}(a)$ and $E_{B}(b)$, associated with the outcomes $A=a$ and $B=b$. The joint probability associated with the state $\ket{\psi_\mathrm{kin}}\in\mathcal{H}_\mathrm{kin}$ is given by ${\mathrm{Pr}[A=a,B=b]=}{\bra{\psi_\mathrm{kin}}E_{A}(a)\otimes E_{B}(b)\ket{\psi_\mathrm{kin}}}$. This is normalised via the normalisation of the kinematical states. Considering physical states, on the other hand, in particular the quantity $\bra{\psi_\mathrm{phys}}E_{A}(a)\otimes E_{B}(b)\ket{\psi_\mathrm{phys}}$, we see that ${\sum_{a,b}\bra{\psi_\mathrm{phys}}E_{A}(a)\otimes E_{B}(b)\ket{\psi_\mathrm{phys}} = \braket{\psi_\mathrm{phys}}{\psi_\mathrm{phys}}_\mathrm{kin}}$, which diverges e.g.~when $\hat{C}$ has a purely continuous spectrum~\cite{ashtekar1994algebraic}, as is the case in the arrival-time problem considered here. This is why the physical inner product is defined as in Eq.~\eqref{eq:abstract_phys_IP}, not as the kinematical one~\cite{hartle1997comparing}.

The quantity $\bra{\psi_\mathrm{phys}}E_{A}(a)\otimes E_{B}(b)\ket{\psi_\mathrm{phys}}$ therefore cannot be interpreted as a joint probability distribution in the non-normalisable case. The same argument applies for operators of the form e.g.~$E_{A}(a)\otimes \mathds{1}_{B}$, likewise forbidding the interpretation of $\bra{\psi_\mathrm{phys}}E_{A}(a)\otimes \mathds{1}_{B}\ket{\psi_\mathrm{phys}}$ as a marginal probability. The arrival-time proposal in~\cite{MacconeSacha2020,roncallo2023does,pintocouto2026first} solves this issue by introducing a regularisation parameter, whose role and suitability is debated in~\cite{cavendish2024physical,maccone2025quantum}. 

To illustrate this in the context of the Page-Wootters formalism, let us consider the case where both observables correspond to clock time. In particular, let $E_{A}(t_{a})=\ketbra{t_{a}}{t_{a}}_{A}$ be the element of the covariant POVM associated with clock $A$ reading $t_{a}$, let $\ket{\psi_{B|A}(t_{a})}$ be the Page-Wootters state of $B$ relative to $A$ associated with the physical state $\ket{\psi_\mathrm{phys}}$, and let us furthermore define the corresponding quantities with respect to $B$. According to the conditional probability interpretation, we then have
\begin{equation} \label{eTwoClockEG}
\begin{aligned}
    \mathrm{Pr}[t_{a}|t_{b}]&\equiv \vert\braket{t_{b}}{\psi_{B|A}(t_{a})}\vert^{2} \\
    &= \bra{\psi_\mathrm{phys}}\big(\ketbra{t_{a}}{t_{a}}_{A}\otimes \ketbra{t_{b}}{t_{b}}_{B}\big)\ket{\psi_\mathrm{phys}} \\
    &= \vert\braket{{t_{a}}}{\psi_{A|B}(t_{b})}\vert^{2} \equiv \mathrm{Pr}[t_{b}|t_{a}]  ,
\end{aligned}
\end{equation}
and integrating over either $t_{a}$ or over $t_{b}$ gives unity, but an integral over both $t_{a}$ and $t_{b}$ diverges. The choice of which conditional probability one attributes to Eqs.~\eqref{eTwoClockEG} is then arbitrary, but neither can be understood in terms of joint and marginal distributions. Both conditional distributions are equal to what in kinematical terms would be the joint distribution, i.e,~the second line in~\eqref{eTwoClockEG}. It is common to divide this term by, e.g., $\bra{\psi_\mathrm{phys}}\big(\ketbra{t_{a}}{t_{a}}_{A}\otimes \mathds{1}_{B}\big)\ket{\psi_\mathrm{phys}}$ to obtain something which better resembles a conditional probability in kinematical terms, but this resemblance is potentially misleading, as ${\bra{\psi_\mathrm{phys}}\big(\ketbra{t_{a}}{t_{a}}_{A}\otimes \mathds{1}_{B}\big)\ket{\psi_\mathrm{phys}}=1}$ $\forall\, t_{a}$, and therefore cannot be interpreted as a marginal probability.

Let us finally return to the case of a free particle, and the notation of Sec.~\ref{secFreeParticlePage-Wootters}. The Page-Wootters conditional probability distribution for the particle's position $x$ at time $t$ is given by
\begin{equation} \label{eXgivenT}
\begin{aligned}
    \mathrm{Pr}[x|t]&\equiv \bra{\psi_{S|C}(t)}\ketbra{x}{x}_{S}\ket{\psi_{S|C}(t)} \\
    &= \frac{1}{2\pi} \bra{\psi_\mathrm{phys}}\big(\ketbra{t}{t}_{C}\otimes \ketbra{x}{x}_{S}\big)\ket{\psi_\mathrm{phys}}.
\end{aligned}
\end{equation}
Again, this has the appearance of a joint probability in the kinematical space (as do all probabilities obtained from $\mathcal{H}_{\mathrm{S|C}}$). Now, unlike in the two-clock example above, we do not have the freedom to interpret Eq.~\eqref{eXgivenT} as $\mathrm{Pr}[t|x]$, i.e. an arrival time probability. Integrating over $t$ gives the norm of the output of the ``naive'' reduction map $\tilde{\mathcal{R}}(x_{0})$ discussed in Sec.~\ref{sConditionPosition}. As noted earlier, this is neither unity nor normalisable in a manner independent of $\ket{\psi_\mathrm{phys}}$.

Considering now the relational arrival time distribution in Sec.~\ref{sToADistibution}, one finds anomalous properties with respect to kinematical structures. In particular, noting that the states  $\ket{x_{0},\sigma}_{S}$ in Eq.~\eqref{ePositionReductonStates} can be written $\ket{x_{0},\sigma}_{S}=\sqrt{\hat{p}/m}\,\Pi^{\sigma}_{S}\ket{x_{0}}_{S}$, where $\Pi^{\sigma}_{S}$ is a projector onto the $\mathrm{sgn}(p)= \sigma$ subspace in $\mathcal{H}_{S}$, we see that $\sum_{\sigma}\int_\mathds{R}dx\,\ketbra{x,\sigma}{x,\sigma}_{S}\not\propto\mathds{1}_{S}$. The states $\lbrace\ket{x,\sigma}_{S}\rbrace_{x}$ therefore do not correspond to the outcomes of some marginal kinematical observable. However, as we have discussed above, there is no meaningful notion of a marginal probability in the physical Hilbert space, and as we saw in Eq.~\eqref{eXgivenT}, even when an operator (in that case, $\ketbra{x}{x}_{S}$) can be associated with a marginal probability in the kinematical space, it cannot necessarily be used to construct a meaningful relational description. Furthermore, as we showed in Sec.~\ref{sConditionPosition}, the reduction maps constructed from $\ket{x_{0},\sigma}_{S}$ give the only possible (up to a phase freedom) relational description of the clock that satisfies the appropriate spatial translation property.

\section{Discussion} \label{sDisc}

We studied the ToA problem for a free particle based on a relational description of the state of the clock relative to the particle's position. The resulting arrival-time distribution in Eq.~\eqref{eToADistributionInit} is the consequence of three conditions which we imposed: 1)~that the Hamiltonian constraint be respected (i.e.~that the relational clock state be derived from $\mathcal{H}_{\mathrm{phys}}$); 2)~that the parameter encoding the particle's position translate appropriately under a translation of the spatial coordinate; 3)~that clock time be covariant with respect to the clock Hamiltonian (the generalisation of canonical conjugacy). Possible extensions of this work include moving beyond free-particle dynamics and our conditioning on a single spatial point. It may be of interest to study external potentials, for example to adapt the approach here to the related problem of tunnelling times~\cite{hauge1989tunneling}, as well as interactions, bearing in mind that clock-system interactions can introduce additional subtleties in relational dynamics~\cite{smith2019quantizing,paiva2022flow,paiva2022non,hausmann2025measurement,rijavec2025conditions}.

Since Eq.~\eqref{eToADistributionInit} coincides with the Kijowski distribution~\cite{kijowski1974time}, it inherits some of the appeal and the criticisms of that approach. For example, it does not incorporate detector dynamics, finite resolution, or interaction effects. However, in contrast with the axiomatic separation of left- and right-moving momentum components in~\cite{kijowski1974time,delgado1997arrival,egusquiza2002standard}, we obtain this separation as a consequence of the constraint, as cross-sector terms are suppressed by the direct–sum structure of $\mathcal{H}_{\mathrm{phys}}$ and the associated superselection rule.\footnote{One may ask how this is compatible with the usual description of a free particle, where there is no restriction on coherence between positive and negative momenta. Somewhat counter-intuitively, the Page-Wootters formalism nonetheless permits such coherences when describing the system relative to the clock, while still respecting the superselection rule at the level of the physical Hilbert space. In particular, any observable on the relational space $\mathcal{H}_{S|C}$ which encodes phase information between positive- and negative-momentum contributions to the particle's wavefunction, is equivalent to an observable on the physical Hilbert space which satisfies the superselection rule there, as can be seen directly from Eq.~(43) of~\cite{hohn2021trinity}.}
One consequence of this sector-wise separation is the prediction that no interference can be observed between counter-propagating wavepackets~\cite{mielnik2011time,roncallo2023does}. Note that an experimental observation of such an interference phenomenon in arrival times would falsify the model explored here, in contrast with usual studies of the Page-Wootters approach, whose role has so far been formal.

A natural question arises here: could a different relational choice within the Page-Wootters framework lead to a different arrival-time distribution, or is the Kijowski distribution inherent to this approach? This is possible only under a modification of at least one of the three conditions described above. Even if an appropriate motivation can be found to break the second and/or third conditions, namely spatial translation covariance and energy-time conjugacy, the resulting modified distribution will still necessarily separate momentum sectors, maintaining the non-interference of counter-propagating wavepackets.

As discussed in Sec.~\ref{sec:necessity_of_conditional_observables}, the reduction map in Eq.~\eqref{ePositionReductonStates} is not associated with a normalised kinematical observable. It is, however, derived from the three conditions above, and so an attempt to restore a marginal interpretation at the kinematical level by replacing Eq.~\eqref{ePositionReductonStates} with states associated with a (kinematically) normalised observable, will either violate one of these conditions, or will somehow reduce to Eq.~\eqref{ePositionReductonStates} when restricted via the constraint. While we have not ruled out the possibility of the latter case, we have been unable to find an example of it.

In Sec.~\ref{secFreeParticlePage-Wootters}, we found that describing a free particle within the Page-Wootters formalism requires a clock with continuous spectrum, implying that clock times range over the entire real line. This contrasts with the proposal in~\cite{MacconeSacha2020,roncallo2023does,pintocouto2026first}, 
where the divergence of the conditional probability is regularised by restricting the clock times to a bounded interval~$\mathbb{T}=[-T/2,T/2]$ for some $T>0$. In the present formalism, however, such a restriction is equivalent to using a periodic clock~\cite{chataignier2026relational}, necessitating a discrete spectrum of $\hat{H}_C$, which is incompatible with a relational description of a free particle. One may ask whether these two perspectives can be reconciled, for example for a period $T$ much larger than some characteristic timescale of the scenario in question. Even if this approximates the free-particle regime sufficiently well, it still leads to a different arrival-time distribution. For a periodic clock and a free particle, the physical Hilbert space maintains the separation into left and right-moving sectors described in Sec.~\ref{secFreeParticlePage-Wootters} (see Example~9 in~\cite{chataignier2026relational}), so the absence of interference between these sectors persists, unlike in~\cite{roncallo2023does}. Consequently, the two approaches represent physically distinct proposals.

Finally, let us compare the approach adopted here with other examples of arrival-time distributions in a relational context. The inequivalence of the approach adopted here with the conditional probability approach in~\cite{MacconeSacha2020,roncallo2023does,pintocouto2026first} was discussed above. The proposal in~\cite{gambini2022solution}, likewise based on conditional probabilities, gives a different prediction for arrival times than the one outlined here; this can be seen by considering the example of a nonideal, but unbounded below Hamiltonian for a clock (e.g. $\mathrm{spec}(\hat{H}_{C})=\mathds{R}_{-}$), which gives reversible behaviour in our case, and irreversible behaviour in~\cite{gambini2022solution}. Finally, it is interesting to note that the Kijowski distribution has also been derived in the relationally-motivated extension of quantum mechanics in~\cite{dias2017space}, where conditional probabilities are constructed with respect to kinematical structures.

\section{Conclusion}
\label{sec:conclusion}

We have constructed a ToA probability distribution within the Page-Wootters framework, which treats time as a relational quantity emerging from clock-system correlations under a global Hamiltonian constraint. By demanding that the approach be grounded in the \emph{physical} Hilbert space, we find a distinct solution from other relational approaches to the problem. Conditioning on the particle being at $x_0$ via a sector-adapted reduction map yielded a well-defined clock state relative to the particle's position, and the covariant time POVM thereon gave a positive, normalised probability distribution for the arrival time. This distribution coincides with a well-known approach in the literature, though here the momentum-branch restriction arises as a consequence of the Hamiltonian constraint, rather than by assumption. However, our approach conflicts with the prevailing interpretation of the Page-Wootters formalism in terms of conditional probabilities, suggesting that this interpretation might not be able to coexist with the spatial translation and time covariance properties of the arrival-time distribution. 

Our work provides a novel, relational approach to the ToA problem on the one hand, and a practical situation in which to examine the Page-Wootters framework on the other. In the latter regard, the non-interference feature of the ToA distribution recovered here presents an example of empirical falsifiability in an otherwise purely formal framework.

\section*{Acknowledgements}

The authors thank Simone Roncallo for helpful discussions. N. H. acknowledges support from ERC-2021-COG 101043705 ``Cocoquest''. This research was funded in whole or in part by the Austrian Science Fund (FWF) 10.55776/I6949, 10.55776/COE1 and the European Union – NextGenerationEU.

\bibliography{biblio}
\onecolumngrid

\appendix

\section{General Page-Wootters reduction for systems with quadratic Hamiltonians} \label{sSectorwiseReduction}
For a quadratic-momentum clock Hamiltonian, i.e.~$\hat{H}_{C}\propto\hat{p}^{2}$, one constructs covariant time observables, and thus Page-Wootters reduction maps, per momentum sector~\cite{hohn2021equivalence}. Here we consider general Page-Wootters-style reduction maps to show that, for such Hamiltonians, any reduction map must likewise be defined sector-wise in order to preserve the physical inner product. We then show that this strongly constrains the form of the reduction maps on each sector, which in turn allows the derivation of the specific example in Sec.~\ref{sConditionPosition}. The proof is given for the specific form of Hamiltonian $\frac{\hat{p}^{2}}{2m}$ considered in the main text, but can be readily generalised to arbitrary quadratic Hamiltonians.

First recall the direct-sum decomposition of the physical Hilbert space into $\sigma$-sectors, i.e. $\mathcal{H}_\mathrm{phys}=\bigoplus_{\sigma = \pm}\mathcal{H}_\mathrm{phys}^{\sigma}$, and
that we can write an arbitrary physical state as
\begin{equation}\label{eArbPhysState}
\ket{\psi_\text{phys}}
= \quad \sum_{\sigma=\pm} \int_{\sigma_{C|S}} d\varepsilon \, \sigma \sqrt{-\frac{m}{2\varepsilon}} \, \psi_{\text{kin}}(\varepsilon, p_{\sigma}(\varepsilon) ) \, \ket{\varepsilon}_C  \ket{p_{\sigma}(\varepsilon)}_{S} 
\end{equation}
with $p_{\sigma}(\varepsilon)\coloneqq\sigma\sqrt{-2m\varepsilon}$, $\sigma_{C|S}\coloneqq\mathrm{spec}(\hat{H}_{C})\cap\mathrm{spec}(-\hat{H}_{S})$ and where $\psi_{\text{kin}}(\varepsilon, p)$ is the wavefunction of any member of the family of kinematical states yielding $\ket{\psi_{\text{phys}}}$ under the projector $\delta(\hat{C})$ defined in Eq.~\eqref{eq:phys_projector} in the main text (see Sec.~\ref{sec:Page-Wootters}). We will now consider an arbitrary Page-Wootters ``conditioning'' with respect to some property of $S$, encoded in a family of states $\lbrace\ket{y}_{S}\in\mathcal{H}_{S}\rbrace_{y}$, parameterised by $y$. Specifically, we consider the family of maps $\lbrace\mathcal{R}(y):\mathcal{H}_\mathrm{phys}\to\mathcal{H}_\mathrm{S\vert C}\rbrace_{y}$, where $\mathcal{H}_{C\vert S}\subseteq\mathcal{H}_{C}$, of the form $\mathcal{R}(y)\coloneqq \mathds{1}_{C}\otimes \bra{y}_{S}$. The result is to be interpreted as the reduced state of $C$, given that the system $S$ is in the state $\ket{y}_{S}$. Let us write this state in the momentum basis, i.e.
\begin{equation} \label{eArbitraryReduction}
\bra{y}_{S} \equiv \int_\mathds{R} dp \, r(y,p) \bra{p}_{S} ,
\end{equation}
for some function $r(y,p)$, which fully determines the reduction map. When applied to a physical state as in Eq.~\eqref{eArbPhysState}, one obtains the following reduced ``states'' on $\mathcal{H}_{C\vert S}$:
\begin{align}
    \ket{\psi_{C\vert S}(y)} &\coloneqq \mathcal{R}(y) \ket{\psi_\mathrm{phys}} \\
    &= \sum_{\sigma=\pm} \int_{\sigma_{C|S}} d\varepsilon \, \sigma \sqrt{-\frac{m}{2\varepsilon}} \, \psi_{\text{kin}}(\varepsilon, p_{\sigma}(\varepsilon) ) \, r(y,p_{\sigma}(\varepsilon)) \, \ket{\varepsilon}_C . 
\end{align}
Noting that for each $\varepsilon$ in the integral above, we have a sum of a $\psi_{\text{kin}}(\varepsilon, p_{+}(\varepsilon) )$ term and a $\psi_{\text{kin}}(\varepsilon, p_{-}(\varepsilon))$ term, we see that $\mathcal{R}(y)$ is many-to-one, and therefore not invertible. For example, the physical state corresponding to $\psi_{\text{kin}}(\varepsilon, p_{\sigma}(\varepsilon) )$, gives the same result under $\mathcal{R}(y)$ as the physical state corresponding to $-\psi_{\text{kin}}(\varepsilon, -p_{\sigma}(\varepsilon) )$ does. Thus the map can only be invertible on one momentum sector $\mathcal{H}_\mathrm{phys}^{\sigma}$.

In addition, considering the inner product between two states in $\mathcal{H}_{C\vert S}$:
\begin{align} \label{eRedInProd}
    \braket{\phi_{C\vert S}(y)}{\psi_{C\vert S}(y)} & =  \int_{\sigma_{C|S}} d\varepsilon \, \left( \sum_{\sigma=\pm} \sigma \sqrt{-\frac{m}{2\varepsilon}} \, \phi_{\text{kin}}(\varepsilon, p_{\sigma}(\varepsilon) ) \,r(y,p_{\sigma}(\varepsilon)) \right) \left( \sum_{{\sigma}'=\pm} {\sigma}' \sqrt{-\frac{m}{2\varepsilon}} \, \psi_{\text{kin}}(\varepsilon, p_{{\sigma}'}(\varepsilon) ) \, r(y,p_{\sigma}(\varepsilon)) \right)^{*}
\end{align}
and comparing this with the physical inner product:
\begin{equation}
\braket{\phi_{\mathrm{phys}}}{\psi_{\mathrm{phys}}}_{\mathrm{phys}}
= \int_{\sigma_{C\vert S}} d\varepsilon\;
   \sqrt{-\frac{m}{2\varepsilon}}\;
   \sum_{\sigma=\pm}
   \phi_{\mathrm{kin}}^{*}\!\bigl(\varepsilon, p_\sigma(\varepsilon)\bigr)\,
   \psi_{\mathrm{kin}}\!\bigl(\varepsilon, p_\sigma(\varepsilon)\bigr)
\end{equation}
we see that the preservation of the inner product, i.e. the condition
\begin{equation} \label{eInProdPreserve}
    \braket{\phi_{C\vert S}(y)}{\psi_{C\vert S}(y)}  \overset{!}{=} \braket{\phi_{\mathrm{phys}}}{\psi_{\mathrm{phys}}}_{\mathrm{phys}} \quad \forall \quad \ket{\psi_{\mathrm{phys}}},\ket{\phi_{\mathrm{phys}}}\in\mathrm{supp}\left( \mathcal{R}(y) \right),
\end{equation}
can only be satisfied on one momentum sector $\mathcal{H}_\mathrm{phys}^{\sigma}$ of the physical Hilbert space. Specifically, on $\mathcal{H}_{+}$ one must have that the $r(y,p)$ in Eq.~\eqref{eArbitraryReduction} satisfy $r(y,p)=0$ $\forall p<0$, or on $\mathcal{H}_{-}$ one must have  $r(y,p)=0$ $\forall p>0$. From Eq.~\eqref{eRedInProd}, we see that a violation of this sector-wise reduction results in the reduced space $\mathcal{H}_{C\vert S}$ encoding information about coherences between superselection sectors of $\mathcal{H}_\mathrm{phys}$, which is forbidden by the constraint.

One must therefore, as in~\cite{hohn2021equivalence}, define a separate reduction map for each sector, or, writing this in terms of states in $\mathcal{H}_{S}$:
\begin{equation} \label{eSectorReduction}
	\ket{y,\sigma}_{S} \coloneqq \int_\mathds{R} dp \, r_{\sigma}(y,p) \ket{p}_{S} ,
\end{equation}
where $r_{\sigma}(y,p)=0$ for $\sigma p <0$, giving a (sector-wise) reduction map $\mathcal{R}_{\sigma}(y)\coloneqq \mathds{1}_{C}\otimes \bra{y,\sigma}_{S}$ from $\mathcal{H}_\mathrm{phys}^{\sigma}$ to a reduced clock space $\mathcal{H}^{\sigma}_{C\vert S}\subseteq\mathcal{H}_{C}$. The condition in Eq.~\eqref{eInProdPreserve} then becomes the condition that $\mathcal{R}_{\sigma}(y):\mathcal{H}_\mathrm{phys}^{\sigma}\to\mathcal{H}^{\sigma}_{C\vert S}$ is an isometry. Then, denoting $\ket{\psi^{\sigma}_{C\vert S}(y)}\coloneqq \mathcal{R}_{\sigma}(y)\ket{\psi_\text{phys}}$, one can demand that
\begin{equation}
	\braket{\psi_{\mathrm{phys}}}{\psi_{\mathrm{phys}}}_{\mathrm{phys}} \overset{!}{=}\sum_{\sigma=\pm}\braket{\psi^{\sigma}_{C\vert S}(y)}{\psi^{\sigma}_{C\vert S}(y)} ,
\end{equation}
which determines the form of $r_{\sigma}(y,p)$. One thus finds that $\mathcal{R}_{\sigma}(y)$ is an isometry if and only if
\begin{equation} \label{eGeneralSigmawiseReduction}
	\ket{y,\sigma}_{S} = \int_\mathds{R} dp \, \Theta(\sigma p) \,\sqrt{\frac{\sigma p}{m}} e^{i \varphi_{\sigma}(y,p)} \ket{p}_{S},
\end{equation}
where $\Theta(p)$ is the Heaviside function, and $\varphi_{\sigma}(y,p)$ is some real function of $p$, whose form entirely determines the reduction map. As an aside, note that this also ensures that $\mathcal{H}^{+}_{C\vert S}\simeq\mathcal{H}^{-}_{C\vert S}$. Identifying $y$ with a clock time $t$, we see that with the choice $\varphi_{\sigma}(t,p)=-\frac{p^{2}}{2m}t$ in Eq.~\eqref{eGeneralSigmawiseReduction}, one recovers the case of a time observable covariant with respect to $\hat{H}_{S}$, cf.~\cite{hohn2021equivalence}.

To find the inverse map, $\mathcal{R}_{\sigma}(y)^{-1}$, note first that
\begin{equation}
    \delta (\hat{C}) \big( \mathds{1}_{C}\otimes\ketbra{y,\sigma}{y,\sigma}_{S} \big) = \int_{\sigma_{C|S}} d\varepsilon \, \int_{\mathds{R}}dp\, \sqrt{-\frac{\left\vert p_{\sigma}(\varepsilon) p\right\vert}{2 m \varepsilon}}   \Theta (\sigma p) e^{i \big[ \varphi_{\sigma}(y,p_{\sigma}(\varepsilon)) - \varphi_{\sigma}(y,p) \big]} \ketbra{\varepsilon}{\varepsilon}_{C}\otimes\ketbra{p_{\sigma}(\varepsilon)}{p}_{S} ,
\end{equation}
which follows from writing $\mathds{1}_{C}$ in the $\ket{\varepsilon}_{C}$ basis, and applying Eqs.~\eqref{eSumSeparationProjector} and~\eqref{eGeneralSigmawiseReduction}. One then finds that
\begin{equation}
    \delta (\hat{C}) \big( \mathds{1}_{C}\otimes\ketbra{y,\sigma}{y,\sigma}_{S} \big) \ket{\varepsilon}_{C}\ket{p_{\sigma'}(\varepsilon)}_{S} = \delta_{\sigma \sigma'}\ket{\varepsilon}_{C}\ket{p_{\sigma'}(\varepsilon)}_{S} ,
\end{equation}
which in turn implies that 
\begin{equation}
    \delta (\hat{C}) \big( \mathds{1}_{C}\otimes\ketbra{y,\sigma}{y,\sigma}_{S} \big)\ket{\psi_\mathrm{phys}}=\ket{\psi_\mathrm{phys}^{\sigma}} \quad \forall \ket{\psi_\mathrm{phys}}\in\mathcal{H}_\mathrm{phys} ,
\end{equation}
where $\ket{\psi_\mathrm{phys}^{\sigma}}$ is the projection of $\ket{\psi_\mathrm{phys}}$ onto $\mathcal{H}_\mathrm{phys}^{\sigma}$. Now, noting that $\mathcal{R}_{\sigma}(y)$ is only invertible on $\ket{\psi_\mathrm{phys}^{\sigma}}$, we see that there the inverse is
\begin{equation}
    \mathcal{R}_{\sigma}(y)^{-1} = \delta (\hat{C}) \big( \mathds{1}_{C}\otimes\ket{y,\sigma}_{S} \big) ,
\end{equation}
as in Eq.~(42) in~\cite{hohn2021equivalence}. Thus the original state in $\mathcal{H}_\mathrm{phys}$ can be obtained by summing the inverses on each sector's reduced state individually, i.e. $\sum_{\sigma}\mathcal{R}_{\sigma}(y)^{-1}\mathcal{R}_{\sigma}(y)=\mathds{1}_\mathrm{phys}$.

\end{document}